\documentclass[prl,aps,twocolumn,superscriptaddress]{revtex4}
\usepackage{graphicx}
\usepackage{amsmath}
\usepackage{bm}

\begin{document}

\title{A systematic correlation between two-dimensional flow topology and the
abstract statistics of turbulence}

\author{W. Brent Daniel}
    \email{wdaniel@lanl.gov}
    \affiliation{Department of Physics, The Ohio State University,
    Columbus, OH 43210}
    \affiliation{Condensed Matter and Thermal Physics Group
    and the Center for Nonlinear Studies, Los Alamos National
    Laboratory, Los Alamos NM, 87545}
\author{Maarten A. Rutgers}
    \affiliation{Department of Physics, The Ohio State University,
    Columbus, OH 43210}

\date{\today}

\begin{abstract}
Velocity differences in the direct enstrophy cascade of
two-dimensional turbulence are correlated with the underlying flow
topology. The statistics of the transverse and longitudinal
velocity differences are found to be governed by different
structures. The wings of the transverse distribution are dominated
by strong vortex centers, whereas, the tails of the longitudinal
differences are dominated by saddles. Viewed in the framework of
earlier theoretical work this result suggests that the transfer of
enstrophy to smaller scales is accomplished in regions of the flow
dominated by saddles.
\end{abstract}

\maketitle

Two-dimensional (2D) turbulence is a fascinating problem with
relevance in areas as wide-ranging as the dynamics of energy
transfer in atmospheric and geophysical flows
\cite{McWilliams94b,Shivamoggi98a} to the basic statistical
mechanics of interacting vortices
\cite{Couder86a,Benzi92a,Dritschel95a}. Three decades of
theoretical and numerical work starting from the seminal ideas of
Kraichnan \cite{Kraichnan67a,Kraichnan75a} and Batchelor
\cite{Batchelor69a} have provided a picture of 2D turbulence based
upon the scaling laws of abstract statistical quantities. This
description remains strikingly incomplete. For example, there is
still no clear physical understanding of the mechanisms by which
energy and enstrophy are transferred between different length
scales in a turbulent flow; nor is there a conclusive picture of
how the intense coherent structures that dominate the statistics
are formed and evolve.

The challenge is to establish a connection between the statistical
measures of turbulence and the physical dynamics of the turbulent
flow field. In the current work, we demonstrate that by
considering correlations between local flow topology
\cite{Okubo70a,Weiss91a,Helman91a,Vorobieff98a,Rivera01a} and
velocity difference probability distributions (PDFs)
\cite{Frisch95a,Sreenivasan96a,Eyink96a,Smith97a,Paret98a,Vorobieff99a},
we can make this important connection. The wings of the
distributions of the transverse and longitudinal velocity
differences are found to be associated with very different
structures: vortex centers and saddles, respectively. As a
consequence, it will turn out that the transfer of enstrophy must
be accomplished near saddle points. This transfer is the result of
a topological asymmetry in the turbulent flow manifest in the
longitudinal velocity difference PDFs. Furthermore, since the
wings of the longitudinal velocity differences are dominated by
strong saddles, a complete understanding of intermittency in 2D
must include a motivation for the formation of these structures.
This powerful technique can be extended to other statistical
quantities to infer correlations between flow topology and
turbulent dynamics.

\begin{figure}
\includegraphics[width=3.2in]{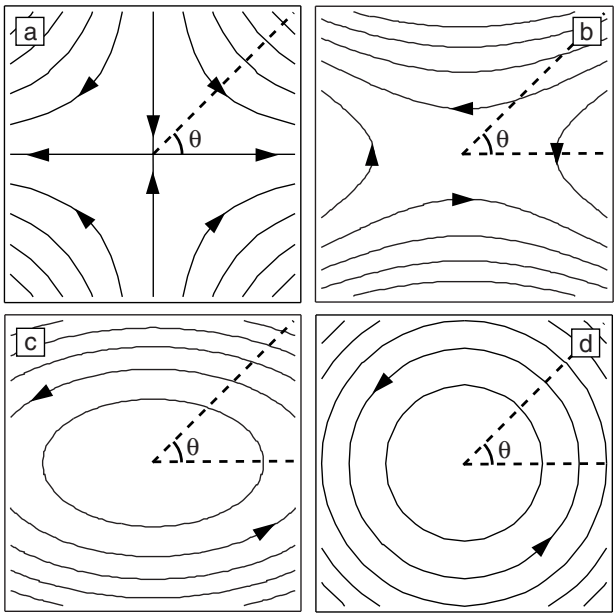}
\caption{Fundamental topological structures of 2D incompressible
flow. The Jacobian determinant, $\Lambda = (1/4)(\omega^2 -
\sigma^2)$, in (a) through (d) reflects a smooth transition from a
dominant strain rate,  $\sigma$, to a dominate vorticity,
$\omega$.} \label{fig:structures}
\end{figure}

\begin{figure}
\includegraphics[width=3.2in]{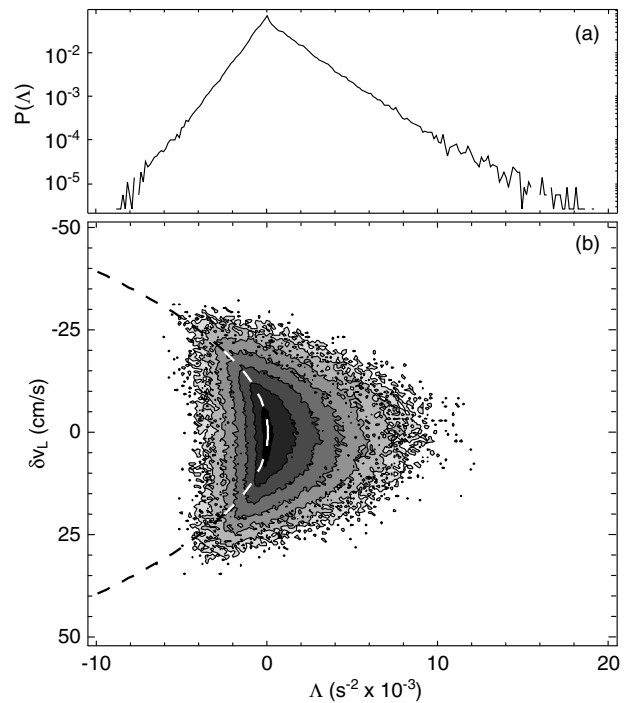}
\caption{(a) The asymmetric distribution of the Jacobian
determinant, $P(\tilde{\Lambda})$. (b) Multivariate probability
distribution, $P(\tilde{\Lambda},\delta v_L(r=0.4 \; {\rm cm}))$,
indicating the likelihood that a given longitudinal velocity
difference will be found in a region of the flow with topology
described by $\tilde{\Lambda}$. Dashed line indicates the velocity
difference of maximum likelihood. Note that the direction of
increasing velocity difference along the vertical axis is
non-standard. Shading and contours represent the log of the
probability.} \label{fig:dvJLong}
\end{figure}

Local flow topology is characterized by the four first-order
derivatives in the expansion of the vector velocity,
\begin{equation}
    \hat{m} =
    \left(
    \begin{array}{cc}
        \partial_x v_x    & \partial_y v_x \\
        \partial_x v_y    & \partial_y v_y
    \end{array}
    \right).
\end{equation}
The determinant of this Jacobian matrix,
\begin{equation}
\Lambda = \det (\hat{m}) = \frac{1}{4}(\omega^2 - \sigma^2),
\end{equation}
represents a local balance between the vorticity and strain rate.
A continuum of structural possibilities exists as the relative
magnitude of these terms is varied; from a symmetric saddle when
the strain rate dominates; to a linear shear region when the two
are of equal magnitude; to an axisymmetric vortex center when the
enstrophy dominates; see Fig. \ref{fig:structures}a-d. Rivera, Wu,
and Yeung \cite{Rivera01a} showed that the probability
distribution of the Jacobian determinant is non-analytic at
$\Lambda=0$, since centers and saddles are topologically distinct,
and asymmetric, with vortex centers being significantly more
likely than saddles of comparable strength; see Fig.
\ref{fig:dvJLong}a.

A statistical correlation between the local flow topology and the
velocity differences is constructed from data obtained from a 2D
flowing soap film experiment, the configuration of which is
described in Ref. \cite{Rutgers01a} with an effective injection
scale of 2 cm. The Jacobian determinant at each location is
calculated from the matrix, $\hat{m}$, averaged over a disk,
$\Omega$, with center halfway between the two velocity
measurements and radius, $r_\Omega$, equal to $r/2$. That is,
$\tilde{\Lambda} = \det(\hat{M})$, where
\begin{equation}
M^{\alpha \beta} = \frac{\int_\Omega m^{\alpha \beta} \, {\mathbf
dr}}{\int_\Omega {\mathbf dr}}.
\end{equation}
By performing the average in this manner a scale-dependence of the
quantity is maintained, allowing the method to probe different
regions of the enstrophy cascade. Here, however, we will be
reporting results for only one separation, $r=0.4$ cm. This
average over the smaller scales in the flow is permissible in the
enstrophy cascade since enstrophy transfer through a given scale
depends only on larger scales and not on these smaller structures
\cite{Falkovich94a}.

There is some sensitivity to flow inhomogeneities in this
measurement since we are averaging over a macroscopic region of
the flow. The inhomogeneity is characterized by a 25\% variation
in the turbulence intensity across the 2.0 cm wide measurement
area. This variation is due to the particular turbulent forcing
mechanism used (a pair of combs arranged in an inverted wedge
\cite{Rutgers98a}). Nevertheless, over a 0.28 cm wide stripe down
the central region of the channel where we gather statistics the
turbulence intensity varies by less than 1\%. Further, a simple
line average between the two points at which the velocity is
measured---less affected by cross-stream inhomogeneity---produced
the same results.

\begin{figure}
\includegraphics[width=3.2in]{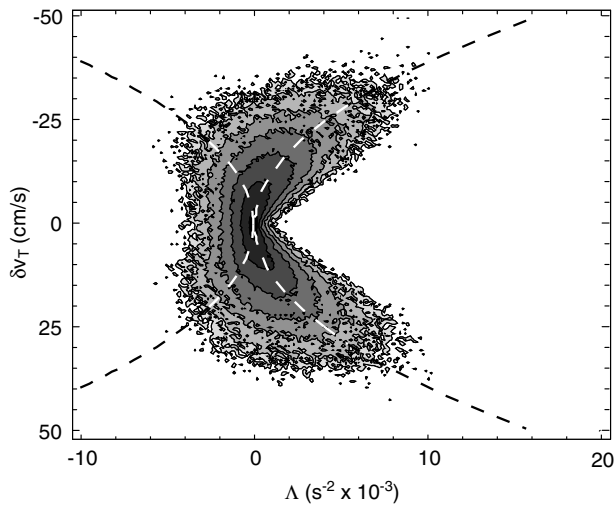}
\caption{Multivariate distribution, $P(\tilde{\Lambda},\delta
v_T(r=0.4\; {\rm cm}))$, indicating the likelihood of finding a
given transverse velocity difference in a region of the flow with
topology described by $\tilde{\Lambda}$. Dashed lines indicate the
velocity difference of maximum likelihood.} \label{fig:dvJTrans}
\end{figure}

A rigorous connection between the abstract velocity difference
statistics and the concrete flow topology is established through
multivariate probability distributions of the form
$P(\tilde{\Lambda},\delta v_i(r))$, where $i$ represents either
the transverse, $T$, or longitudinal component, $L$, of the
velocity difference. The form of these distributions differs
significantly between the longitudinal and transverse velocity
increments as a result of the different symmetries of the saddle
and center; see Figs. \ref{fig:dvJLong} and \ref{fig:dvJTrans}.
The wings of the longitudinal velocity difference PDF are
dominated by strong saddles, whereas, the wings of the transverse
velocity difference are dominated by strong centers. To understand
the reason for this distinction we examine the distribution of the
velocity differences about these two first-order structures in
some detail.

\begin{figure}
\includegraphics[width=3.2in]{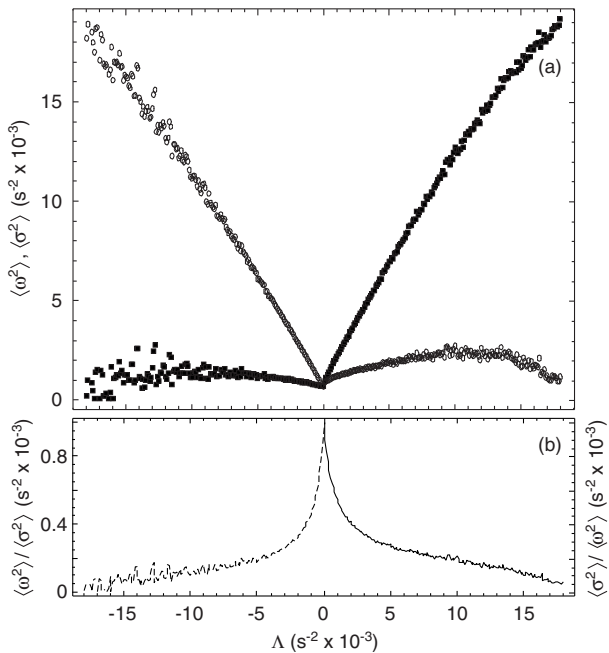}
\caption{(a) The expected values of $\langle \omega^2 \rangle$,
filled squares, and $\langle \sigma^2 \rangle$, open circles. For
positive and negative $\tilde{\Lambda}$ different topologies tend
to be significantly dominant, as is indicated by (b), the ratios
$\langle \omega^2 \rangle / \langle \sigma^2 \rangle$ (dashed
line) and $\langle \sigma^2 \rangle / \langle \omega^2 \rangle$
(solid line) for negative and positive $\tilde{\Lambda}$
respectively.} \label{fig:Expectations}
\end{figure}

We begin with an examination of a saddle point. The symmetry of a
saddle is such that the distribution of the longitudinal and
transverse velocity differences about it is the same. If the
matrix, $\hat{m}$, is parametrized as
\begin{equation}
    \hat{m} =
    \frac{1}{\sqrt{2}}
    \left(
    \begin{array}{cc}
        \lambda   & \Delta + \omega \\
        \Delta - \omega    & -\lambda
    \end{array}
    \right),
\end{equation}
the radial and angular components of the velocity about a saddle
(described by the symmetric part of $\hat{m}$) are given by $v_r =
\left[ \lambda \cos(2\theta) + \Delta \sin(2\theta) \right]r$ and
$v_\theta = \left[ \lambda \sin (2 \theta) - \Delta \cos (2
\theta) \right] r$, respectively. As long as $\sigma^2 =
\lambda^2+\Delta^2$ is held constant, the relative magnitudes of
$\lambda$ and $\Delta$ serve only to vary the spatial orientation
of the saddle. We can, therefore, make the simplifying assumption
that the saddle is described by $\sigma^2 = \lambda^2$. Using
$P(v_r(\theta)) \propto \left[
\partial_\theta v_r(\theta) \right]^{-1}|_{\theta(v_r)}$ and $P(v_\theta(\theta)) \propto \left[
\partial_\theta v_\theta(\theta) \right]^{-1}|_{\theta(v_\theta)}$ in
the longitudinal and transverse cases respectively, we obtain
probability distributions of the form
\begin{equation}
P(\delta v_i (r)|\tilde{\Lambda}) = \frac{2}{\pi \sqrt{r^2
\Lambda^2 - \delta v_i^2}}, \label{eqn:PdvL}
\end{equation}
regardless of which velocity increment is examined. Here we have
let, for example, $\delta v_{\rm L} = 2 v_r$, and replaced the
independent variables $\lambda$ and $\Delta$ by an equivalent pair
expressing the saddle strength, $\sigma^2$, and orientation.
Because of this similarity, it's not surprising that the
appearance of the multivariate distributions, $P(\tilde{\Lambda},
\delta v_T(r))$ and $P(\tilde{\Lambda}, \delta v_L(r))$, are, at
first order, identical for $\tilde{\Lambda} \ll 0$ (where
$\tilde{\Lambda} \simeq -(1/4)\sigma^2$). The velocity differences
of maximum likelihood for a given saddle strength follow the curve
$\delta v_i = \pm \, r \sqrt{-\tilde{\Lambda}}$ for
$\tilde{\Lambda} < 0$, independent of which velocity difference is
being considered (the dashed lines for $\tilde{\Lambda}<0$ in
Figs. \ref{fig:dvJLong}b and \ref{fig:dvJTrans}). Note that the
assumption that the dominant topology takes the form of a saddle
is not valid as $\tilde{\Lambda}$ nears zero. In this regime, Fig.
\ref{fig:Expectations} shows that the expected values of the
squared strain rate and squared vorticity become of comparable
magnitude. The form of the $P(\delta v_i (r)|\tilde{\Lambda})$ is,
therefore, distorted as the topological features themselves are
stretched (recall Fig. \ref{fig:structures}).

Although the distributions of the longitudinal and transverse
velocity differences are identical across a saddle, the
distribution of these two quantities about a vortex center differs
dramatically. Across an axisymmetric vortex center the
longitudinal velocity difference is precisely zero (since the
radial component of the velocity itself is zero). For positive
values of the Jacobian determinant, points in the distribution
$P(\tilde{\Lambda}, \delta v_L(r))$ which lie away from $\delta
v_{\rm L} = 0$ do so either as the result of first order
contributions from the strain rate or through higher order
corrections to the flow field. Since the magnitude of the
longitudinal velocity difference across a vortex center is, thus,
constrained to be small, it is saddle-like regions of the flow
which play the dominant role in the wings of the longitudinal
velocity difference PDFs and, correspondingly, in the higher order
longitudinal structure functions. A clear picture of intermittency
in 2D must, therefore, include an understanding of the formation
of unusually strong saddle points in addition to coherent vortex
centers.

On the other hand, vortex centers do support significant {\it
transverse} velocity differences. In fact, vortex centers with
$\delta v_{\rm T}=0$ cannot exist as a consequence of Stoke's law:
$\oint_C v_\theta r_\Omega \, d\theta = \int_\Omega \omega_z \,
d\Omega$, where $C$ traces the perimeter of the disk, $\Omega$,
and for $\tilde{\Lambda} > 0$, the sign of $v_\theta$ is the same
about the entire perimeter. As the magnitude of the transverse
velocity goes to zero, so must the magnitude of the vorticity. The
shape of the distribution is obtained under the assumption that
for locations in the flow with $\tilde{\Lambda} \gg 0$ the
topology is dominated by the vorticity. Setting $\tilde{\Lambda} =
(1/4)\,\omega^2$, $v_\theta = \omega\, r_\Omega$, $\delta v_{\rm
T} = 2 v_\theta$, and letting $r_\Omega = r/2$, we find that the
transverse velocity difference varies as a function of the local
Jacobian determinant according to $\delta v_{\rm T} = \pm \, r
\sqrt{\tilde{\Lambda}}$ for $\tilde{\Lambda}>0$ (the dashed line
in Fig. \ref{fig:dvJTrans}). The finite spread in the distribution
results both from higher order corrections to the shape of the
vortex centers as well as from the fact that in actuality there
are finite contributions from first-order saddles, that is,
$\langle \sigma^2 \rangle / \langle e_s \rangle$ is of order 0.2
for $\tilde{\Lambda} > 0$.

Because of the larger propensity for strong vortex centers to form
relative to saddle points of comparable magnitude (recall Fig.
\ref{fig:dvJLong}a), the maximum transverse velocity difference
found about centers in the flow is significantly greater than that
found about saddles (compare Fig. \ref{fig:dvJTrans} for
$\tilde{\Lambda}<0$ and $\tilde{\Lambda}>0$). The wings of the
transverse velocity difference PDF in the enstrophy cascade are,
therefore, dominated by contributions from vortex centers. Because
of this clear segregation between the structures which play the
dominant role in the transverse and longitudinal velocity
differences, it would not be surprising if the higher order
moments of these two quantities differed.

These rudimentary observations have certainly not expended the
usefulness of this technique. It is, first of all, interesting
that the ratio $\langle \sigma^2 \rangle / \langle \omega^2
\rangle$ tends toward zero for $\tilde{\Lambda} \gg 1$. This
implies that the strongest vortex centers are nearly axisymmetric,
Fig. \ref{fig:structures}b rather than \ref{fig:structures}d. This
was not seen in the earlier continuously forced experiment of
Rivera {\it et al.}\cite{Rivera01a}. These axisymmetric centers
are the likely predecessors of the coherent structures found in
the latter stages of 2D turbulent decay\cite{Smith97a,Bracco00a}.
They are stable \cite{McWilliams84a,Whitaker94a,Kimura01a},
steady-state solutions of the Euler equation in which there is no
nonlinear transfer of enstrophy, and hence, no enstrophy cascade
\cite{Babiano87a}.

Furthermore, a key feature is still missing from the first order
distribution, $P(\delta v_{\rm L}(r)|\tilde{\Lambda})$. It is a
well know result in turbulence theory that the rate of enstrophy
transfer depends on an odd moment of the longitudinal velocity
difference \cite{Bernard99a}, $S^{(\rm L)}_3(r) = (1/8) \, \eta \,
r^3$, and, hence, on an asymmetry in the distribution of these
velocity differences. This asymmetry is missing in Eq.
(\ref{eqn:PdvL}), where $P(\delta v_{\rm L}|\tilde{\Lambda}) =
P(-\delta v_{\rm L}|\tilde{\Lambda})$. In fact, any asymmetry is
explicitly prohibited by incompressibility in the linear field
approximation. Second order terms, as well, of the form
$(\partial_{xx} v_y) \, x^2$ and $(\partial_{yy} v_x) \, y^2$ are
even under the operations $x \rightarrow -x$ and $y \rightarrow
-y$ so that the contribution to the velocity on opposite sides of
the structure is of the same magnitude and sign. Incompressibility
inextricably links the remaining $\partial_{xy}$ terms. The result
is that no contribution to the longitudinal asymmetry is possible
at order two. It is necessary to go to at least third order to
explain the asymmetry in $P(\delta v_{\rm L})$, where the
$\partial_{xxx} v_x$ and $\partial_{yyy} v_y$ terms---and through
incompressibility the $\partial_{xxy} v_y$ and $\partial_{xyy}
v_x$ terms---result in an asymmetric saddle; that is, one in which
the magnitude of the velocity in the incoming and out-going jets
differs. The longitudinal velocity difference about such a
structure is asymmetric, whereas, positive and negative transverse
velocity increments continue to exist with equal likelihood. A
systematic correlation between these higher order topological
structures and the velocity differences is beyond the limits of
the current data set, but future work will explore these
connections.

We would like to thank Michael Rivera, Robert Ecke, and Michael
Chertkov for many insightful conversations. The present work was
supported by the Petroleum Research Foundation, The Ohio State
University, and the U.S. Department of Energy (W-7405-ENG-36).


\begin{thebibliography}{10}
\expandafter\ifx\csname bibnamefont\endcsname\relax
  \def\bibnamefont#1{#1}\fi
\expandafter\ifx\csname bibfnamefont\endcsname\relax
  \def\bibfnamefont#1{#1}\fi
\expandafter\ifx\csname url\endcsname\relax
  \def\url#1{\texttt{#1}}\fi
\expandafter\ifx\csname
urlprefix\endcsname\relax\def\urlprefix{URL }\fi
\expandafter\ifx\csname bibinfo\endcsname\relax
\def\bibinfo#1#2{#2}\fi \expandafter\ifx\csname
eprint\endcsname\relax \def\eprint#1{#1}\fi

\bibitem{McWilliams94b}
\bibinfo{author}{\bibfnamefont{J.~C.} \bibnamefont{McWilliams}},
  \bibinfo{author}{\bibfnamefont{J.~B.} \bibnamefont{Weiss}}, \bibnamefont{and}
  \bibinfo{author}{\bibfnamefont{I.}~\bibnamefont{Yavneh}},
  \bibinfo{journal}{Science}
  \textbf{\bibinfo{volume}{264}}(\bibinfo{number}{5157}), \bibinfo{pages}{410}
  (\bibinfo{year}{1994}).

\bibitem{Shivamoggi98a}
\bibinfo{author}{\bibfnamefont{B.}~\bibnamefont{Shivamoggi}},
  \bibinfo{journal}{Ann Phys} \textbf{\bibinfo{volume}{270}},
  \bibinfo{pages}{263} (\bibinfo{year}{1998}).

\bibitem{Couder86a}
\bibinfo{author}{\bibfnamefont{Y.}~\bibnamefont{Couder}} \bibnamefont{and}
  \bibinfo{author}{\bibfnamefont{C.}~\bibnamefont{Basdevant}},
  \bibinfo{journal}{J Fluid Mech} \textbf{\bibinfo{volume}{173}},
  \bibinfo{pages}{225} (\bibinfo{year}{1986}).

\bibitem{Benzi92a}
\bibinfo{author}{\bibfnamefont{R.}~\bibnamefont{Benzi}} \bibnamefont{and}
  \bibinfo{author}{\bibfnamefont{M.}~\bibnamefont{Colella}},
  \bibinfo{journal}{Phys Fluids}
  \textbf{\bibinfo{volume}{4}}(\bibinfo{number}{5}), \bibinfo{pages}{1036}
  (\bibinfo{year}{1992}).

\bibitem{Dritschel95a}
\bibinfo{author}{\bibfnamefont{D.}~\bibnamefont{Dritschel}},
  \bibinfo{journal}{J Fluid Mech} \textbf{\bibinfo{volume}{293}},
  \bibinfo{pages}{269} (\bibinfo{year}{1995}).

\bibitem{Kraichnan67a}
\bibinfo{author}{\bibfnamefont{R.}~\bibnamefont{Kraichnan}},
  \bibinfo{journal}{Phys Fluids} \textbf{\bibinfo{volume}{10}},
  \bibinfo{pages}{1417} (\bibinfo{year}{1967}).

\bibitem{Kraichnan75a}
\bibinfo{author}{\bibfnamefont{R.}~\bibnamefont{Kraichnan}},
  \bibinfo{journal}{J Fluid Mech} \textbf{\bibinfo{volume}{67}},
  \bibinfo{pages}{155} (\bibinfo{year}{1975}).

\bibitem{Batchelor69a}
\bibinfo{author}{\bibfnamefont{G.}~\bibnamefont{Batchelor}},
  \bibinfo{journal}{Phys Fluids Suppl II} \textbf{\bibinfo{volume}{12}},
  \bibinfo{pages}{233} (\bibinfo{year}{1969}).

\bibitem{Okubo70a}
\bibinfo{author}{\bibfnamefont{A.}~\bibnamefont{Okubo}},
  \bibinfo{journal}{Deep-Sea Research} \textbf{\bibinfo{volume}{17}},
  \bibinfo{pages}{445} (\bibinfo{year}{1970}).

\bibitem{Weiss91a}
\bibinfo{author}{\bibfnamefont{J.~B.} \bibnamefont{Weiss}},
  \bibinfo{journal}{Physica D} \textbf{\bibinfo{volume}{48}},
  \bibinfo{pages}{273} (\bibinfo{year}{1991}).

\bibitem{Helman91a}
\bibinfo{author}{\bibfnamefont{J.~L.} \bibnamefont{Helman}} \bibnamefont{and}
  \bibinfo{author}{\bibfnamefont{L.}~\bibnamefont{Hesselink}},
  \bibinfo{journal}{IEEE Computer Graphics and Applications}
  \textbf{\bibinfo{volume}{11}}, \bibinfo{pages}{36} (\bibinfo{year}{1991}).

\bibitem{Vorobieff98a}
\bibinfo{author}{\bibfnamefont{P.}~\bibnamefont{Vorobeiff}} \bibnamefont{and}
  \bibinfo{author}{\bibfnamefont{R.~E.} \bibnamefont{Ecke}},
  \bibinfo{journal}{Physica D} \textbf{\bibinfo{volume}{123}},
  \bibinfo{pages}{153} (\bibinfo{year}{1998}).

\bibitem{Rivera01a}
\bibinfo{author}{\bibfnamefont{M.}~\bibnamefont{Rivera}},
  \bibinfo{author}{\bibfnamefont{X.~L.} \bibnamefont{Wu}}, \bibnamefont{and}
  \bibinfo{author}{\bibfnamefont{C.}~\bibnamefont{Yeung}},
  \bibinfo{journal}{Phys Rev Lett}
  \textbf{\bibinfo{volume}{87}}(\bibinfo{number}{4}), \bibinfo{pages}{044501}
  (\bibinfo{year}{2001}).

\bibitem{Sreenivasan96a}
\bibinfo{author}{\bibfnamefont{K.}~\bibnamefont{Sreenivasan}},
  \bibinfo{author}{\bibfnamefont{S.}~\bibnamefont{Vainshtein}},
  \bibinfo{author}{\bibfnamefont{R.}~\bibnamefont{Bhiladvala}},
  \bibinfo{author}{\bibfnamefont{I.}~\bibnamefont{San~Gil}},
  \bibinfo{author}{\bibfnamefont{S.}~\bibnamefont{Chen}}, \bibnamefont{and}
  \bibinfo{author}{\bibfnamefont{N.}~\bibnamefont{Cao}}, \bibinfo{journal}{Phys
  Rev Lett} \textbf{\bibinfo{volume}{77}}, \bibinfo{pages}{1488}
  (\bibinfo{year}{1996}).

\bibitem{Eyink96a}
\bibinfo{author}{\bibfnamefont{G.}~\bibnamefont{Eyink}},
  \bibinfo{journal}{Physica D} \textbf{\bibinfo{volume}{91}},
  \bibinfo{pages}{97} (\bibinfo{year}{1996}).

\bibitem{Smith97a}
\bibinfo{author}{\bibfnamefont{L.}~\bibnamefont{Smith}} \bibnamefont{and}
  \bibinfo{author}{\bibfnamefont{V.}~\bibnamefont{Yakhot}},
  \bibinfo{journal}{Phys Rev E} \textbf{\bibinfo{volume}{55}},
  \bibinfo{pages}{5458} (\bibinfo{year}{1997}).

\bibitem{Paret98a}
\bibinfo{author}{\bibfnamefont{J.}~\bibnamefont{Paret}} \bibnamefont{and}
  \bibinfo{author}{\bibfnamefont{P.}~\bibnamefont{Tabeling}},
  \bibinfo{journal}{Phys Fluids} \textbf{\bibinfo{volume}{10}},
  \bibinfo{pages}{3126} (\bibinfo{year}{1998}).

\bibitem{Vorobieff99a}
\bibinfo{author}{\bibfnamefont{P.}~\bibnamefont{Vorobieff}},
  \bibinfo{author}{\bibfnamefont{M.}~\bibnamefont{Rivera}}, \bibnamefont{and}
  \bibinfo{author}{\bibfnamefont{R.~E.} \bibnamefont{Ecke}},
  \bibinfo{journal}{Phys Fluids} \textbf{\bibinfo{volume}{11}},
  \bibinfo{pages}{2167} (\bibinfo{year}{1999}).

\bibitem{Frisch95a}
\bibinfo{author}{\bibfnamefont{U.}~\bibnamefont{Frisch}},
  \emph{\bibinfo{title}{Turbulence: The Legacy of A. N. Kolmogorov}}
  (\bibinfo{publisher}{Cambridge University Press},
  \bibinfo{address}{Cambridge}, \bibinfo{year}{1995}).

\bibitem{Rutgers01a}
\bibinfo{author}{\bibfnamefont{M.}~\bibnamefont{Rutgers}},
  \bibinfo{author}{\bibfnamefont{X.~L.} \bibnamefont{Wu}}, \bibnamefont{and}
  \bibinfo{author}{\bibfnamefont{W.~B.} \bibnamefont{Daniel}},
  \bibinfo{journal}{Rev Sci Inst}
  \textbf{\bibinfo{volume}{72}}(\bibinfo{number}{7}), \bibinfo{pages}{3025}
  (\bibinfo{year}{2001}).

\bibitem{Falkovich94a}
\bibinfo{author}{\bibfnamefont{G.}~\bibnamefont{Falkovich}} \bibnamefont{and}
  \bibinfo{author}{\bibfnamefont{V.}~\bibnamefont{Lebedev}},
  \bibinfo{journal}{Phys Rev E}
  \textbf{\bibinfo{volume}{50}}(\bibinfo{number}{5}), \bibinfo{pages}{3883}
  (\bibinfo{year}{1994}).

\bibitem{Rutgers98a}
\bibinfo{author}{\bibfnamefont{M.}~\bibnamefont{Rutgers}},
  \bibinfo{journal}{Phys Rev Lett} \textbf{\bibinfo{volume}{81}},
  \bibinfo{pages}{2244} (\bibinfo{year}{1998}).

\bibitem{Bracco00a}
\bibinfo{author}{\bibfnamefont{A.}~\bibnamefont{Bracco}},
  \bibinfo{author}{\bibfnamefont{J.~C.} \bibnamefont{McWilliams}},
  \bibinfo{author}{\bibfnamefont{G.}~\bibnamefont{Murante}},
  \bibinfo{author}{\bibfnamefont{A.}~\bibnamefont{Provenzale}},
  \bibnamefont{and} \bibinfo{author}{\bibfnamefont{J.~B.} \bibnamefont{Weiss}},
  \bibinfo{journal}{Phys Fluids}
  \textbf{\bibinfo{volume}{12}}(\bibinfo{number}{11}), \bibinfo{pages}{2931}
  (\bibinfo{year}{2000}).

\bibitem{McWilliams84a}
\bibinfo{author}{\bibfnamefont{J.}~\bibnamefont{McWilliams}},
  \bibinfo{journal}{J Fluid Mech} \textbf{\bibinfo{volume}{146}},
  \bibinfo{pages}{21} (\bibinfo{year}{1984}).

\bibitem{Whitaker94a}
\bibinfo{author}{\bibfnamefont{N.}~\bibnamefont{Whitaker}} \bibnamefont{and}
  \bibinfo{author}{\bibfnamefont{B.}~\bibnamefont{Turkington}},
  \bibinfo{journal}{Phys Fluids}
  \textbf{\bibinfo{volume}{6}}(\bibinfo{number}{12}), \bibinfo{pages}{3963}
  (\bibinfo{year}{1994}).

\bibitem{Kimura01a}
\bibinfo{author}{\bibfnamefont{Y.}~\bibnamefont{Kimura}} \bibnamefont{and}
  \bibinfo{author}{\bibfnamefont{J.~R.} \bibnamefont{Herring}},
  \bibinfo{journal}{J Fluid Mech} \textbf{\bibinfo{volume}{439}},
  \bibinfo{pages}{43} (\bibinfo{year}{2001}).

\bibitem{Babiano87a}
\bibinfo{author}{\bibfnamefont{A.}~\bibnamefont{Babiano}},
  \bibinfo{author}{\bibfnamefont{C.}~\bibnamefont{Basdevant}},
  \bibinfo{author}{\bibfnamefont{B.}~\bibnamefont{Legras}}, \bibnamefont{and}
  \bibinfo{author}{\bibfnamefont{R.}~\bibnamefont{Sadourny}},
  \bibinfo{journal}{J Fluid Mech} \textbf{\bibinfo{volume}{183}},
  \bibinfo{pages}{379} (\bibinfo{year}{1987}).

\bibitem{Bernard99a}
\bibinfo{author}{\bibfnamefont{D.}~\bibnamefont{Bernard}},
  \bibinfo{journal}{Phys Rev E} \textbf{\bibinfo{volume}{60}},
  \bibinfo{pages}{6184} (\bibinfo{year}{1999}).

\end{thebibliography}
\end{document}